# Influence of Fe, Ni, and Cu Doping on the Photocatalytic Efficiency of ZnS: Implications for Prebiotic Chemistry

Wei Wang

W. Wang (✉)

CCMST, Academy of Fundamental and Interdisciplinary Sciences,

Harbin Institute of Technology, Harbin 150080, China

e-mail: wwang_ol@hit.edu.cn

**Abstract**

The mineral sphalerite (ZnS) is a typical constituent at the periphery of submarine hydrothermal deposits on Earth. It has been frequently suggested to have played an important role in the prebiotic chemistry due to its prominent photocatalytic activity. Nevertheless, the need for $\lambda < 344$ nm UV radiation, which accounts for a very minor part of the energy range of the incoming solar spectrum, limits the application of this semiconductor. In this paper we employed a simple co-precipitation method for the fabrication of Fe, Ni, and Cu-doped ZnS colloids and investigated their activities in the photocatalyzed reduction of fumaric acid. The results show that the photocatalytic activity of pristine ZnS is almost identical with that of 0.1 atom% Fe-doped ZnS, but decreases by doping 0.1 atom% Ni. However, it can be significantly enhanced by doping Cu because this dopant makes the optical absorption edges of ZnS shift from UV band to longer wavelengths. The optimal doping concentration was found to be 0.3 atom%. Even under $\lambda > 450$ nm light irradiation, the photocatalyst $Zn_{1-x}Cu_xS$ can drive the reduction of fumaric acid to produce succinic acid. Given the existence of this doped semiconductor in the hydrothermal vents on early Earth and its capability to utilize both UV and visible light, ZnS might have participated more efficiently than ever estimated in the prebiotic chemical evolution.

**Introduction**

Zinc sulfide (ZnS) is an n-type semiconductor characterized by a wide bandgap of about 3.6 eV. It can absorb UV light with a wavelength shorter than 344 nm to produce photoelectron/hole pairs and catalyze redox reactions (Wang et al. 2012a). The reaction of the photoelectron with an adsorbate leads to the reduction of the adsorbate, while the hole induces oxidation reactions. This property of ZnS could not only make possible the carbon/nitrogen fixation (Eggins et al. 1993; Ranjit et al. 1994)and molecular chain extension processes (Zhang and Martin 2006; Guzman and Martin 2009; Guzman and Martin 2010) to form organic molecules, but also help maintain the redox homeostasis between certain biomolecules (Wang et al. 2012a). The mineral sphalerite (ZnS) is a typical constituent at the periphery of hydrothermal deposits of shallow-water hot springs on early Earth (Edmond et al. 1995; Mulkidjanian 2009). Since the discovery of submarine hydrothermal vents in 1977 (Corliss et al. 1979), the sulphide world therein has been frequently argued as a very potential cradle for the origins of life (Wächtershäuser 1988; Russell and Hall 1997; Wang et al. 2011). If the hydrothermal vents on early Earth were located at a shallow depth no more than 100 m which consisted of the "photic zone" of primitive ocean, sunlight could penetrate the waters of early Earth to trigger prebiotic synthesis on ZnS surfaces (Wang et al. 2012b; Wang et al. 2013).

However, the fact that ZnS absorbs only light at wavelengths shorter than 344 nm makes it seem unattractive to trigger prebiotic redox reactions. Although the solar light reaching the oxygen-deficient early Earth contained a UV component several orders of magnitude stronger than it is today (Mulkidjanian 2009), the dominant part of the solar spectrum is visible light (Jelle 2013). The wide bandgap of ZnS limits its ability for solar energy harvesting. Therefore, ZnS may become a more promising photocatalyst for prebiotic chemistry if it could response to longer wavelengths. Doping foreign elements into a semiconductor is an effective strategy to narrow its bandgap by modifying its electronic band structure (Zaleska 2008). For instance, Cd- and Bi-doped ZnS solid solutions showed significant visible-light photocatalytic activity for water splitting (Yu et al. 2010; Zhang et al. 2011). However, those elements are rare ones on Earth. In hydrothermal vents, Fe, Ni, and Cu are more common. They always occur in the form of pyrite ($FeS_2$), pentlandite ($(Fe,Ni)_9S_8$), chalcopyrite ($CuFeS_2$), etc (Huber and Wächtershäuser 1997; Nakamura et al. 2010). In the early hydrothermal vents, whether during the formation of pristine ZnS or its remobilization and reprecipitation at the seawater interface (Edmond et al. 1995), these metal ions may co-deposit with ZnS, forming new energy levels in the band structure of ZnS and making the latter responsive to longer wavelength light.

To assess this possibility, we prepared Fe, Ni, and Cu-doped ZnS samples and examined their photocatalytic efficiencies in the reduction of fumaric acid to form succinic acid. This reaction was chosen since it is a biochemical pathway of key importance in the reductive tricarboxylic acid cycle (Zhang and Martin 2006).

**Materials and Methods**

Materials

Fumaric acid and succinic acid (> 99%, Sigma) were used without further purification. All other chemicals (Sinopharm, China) were obtained in analytical grade, with the exception of methanol which was of chromatographic grade. Ultrapure water (Millipore) was deoxygenated with high purity argon gas (99.999%) by bubbling it through 1 L water (1 L/min) for 1 h. The deaerated water was used throughout.

Experimental

$Zn_{1-x}M_xS$ colloidal suspensions were prepared under stirring by dropwise adding an aqueous $Na_2S$ solution (0.3 M, 4 mL) into mixed solutions of $ZnSO_4$ and $FeSO_4$, $NiSO_4$ or $CuSO_4$ (total metal ion concentration: 0.1 M, 10 mL). The nominal atomic percentages of doped metal elements (M, M = Fe, Ni, or Cu) to the total metal atoms (Zn + M) were designated as x. The resulting samples were labeled as $Zn_{1-x}M_xS$.

Each precipitate was centrifuged (4000 rpm), washed with 15 mL water thrice and transferred into a 60-ml quartz conical flask. Then the flask was charged with 9 mmol $Na_2SO_3$, 360 µmol fumaric acid and 30 ml of water. The solution was adjusted to pH 8.5 by adding an appropriate amount of 1 M NaOH. The total liquid volume was brought to 36 ml.

Photochemical reactions were conducted at 25 ℃ under argon atmosphere by using a self-devised photochemical reactor with a 500 W mercury-xenon lamp and a group of long wave pass filters. The general methodology has already been reported elsewhere (Wang et al. 2012a)

After irradiation for a selected time period, 1.0 mL of the suspension was withdrawn from the flask and forced through a 0.22 μm microvoid filter film. HPLC was used to determine the concentration of the succinic acid product with a C18 column (Phenomenex, Luna, 5 μm, 250 × 4.6 mm). A mobile phase consisted of 95 % phosphate buffer (25 mM $KH_2PO_4$, pH3.2) and 5 % methanol was used. The elution behaviors were monitored at 215 nm with a flow rate of 1 mL/min.

**Results**

The photocatalytic activities of pristine ZnS and Fe, Ni, Cu-doped ZnS under irradiation of different wavelength lights are compared in Fig. 1. As can be seen from this figure, under irradiation containing UV components (λ > 320 nm and 354 nm) the photocatalytic efficiencies of ZnS decreased when doped by 0.1 atom% ferrous ion and nickel ions. As the amount of doped Fe or Ni was increased the photocatalytic activity continuously decreased (data not shown). In contrast, the yields of succinic acid catalyzed by $Zn_{1-x}Cu_xS$ (x = 0.1%) were raised about 1.8 and 4.4 times as against by using pristine ZnS at λ > 320 nm and λ > 354nm, respectively. Undoped ZnS (white) and Fe or Ni-doped ZnS samples (pale brown and pale yellow, respectively) showed no visible-light (> 400 nm) activity, while a significant visible-light photocatalytic activity was observed for Cu-doped ZnS (gray white). Since CuS alone also had no any photocatalytic activity (data not shown), the observed enhancing effect implied that $Cu^{2+}$ ions were doped in the ZnS lattice by replacing at Zn sites. This dopant can alter the electronic band structure of ZnS and make its optical absorption edges shift from UV band (344 nm) to longer wavelengths. The photocatalytic effect of ZnS at λ > 354 nm was ascribed to the weak transmission of light with wavelengths shorter than 344 nm through the filter (Fig. 2).

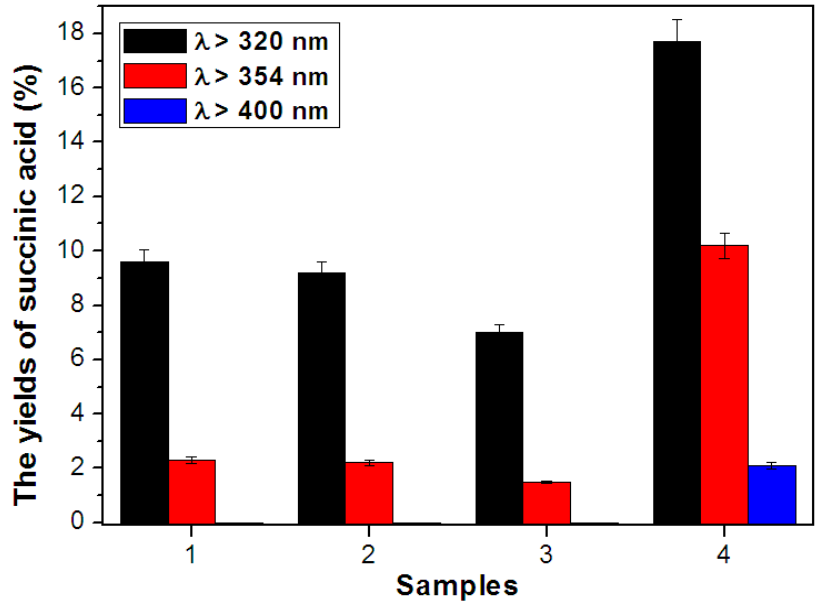

**Fig. 1** The dependence of photocatalytic activities of ZnS for succinic acid evolution over metal ion doping and wavelength. Horizontal-axis: 1. ZnS; 2. $Zn_{1-x}Fe_xS$ (x = 0.1%); 3. $Zn_{1-x}Ni_xS$ (x = 0.1%); 4. $Zn_{1-x}Cu_xS$ (x = 0.1%). Irradiation time: 2 h. The error bars represent standard deviations from three independent experiments.

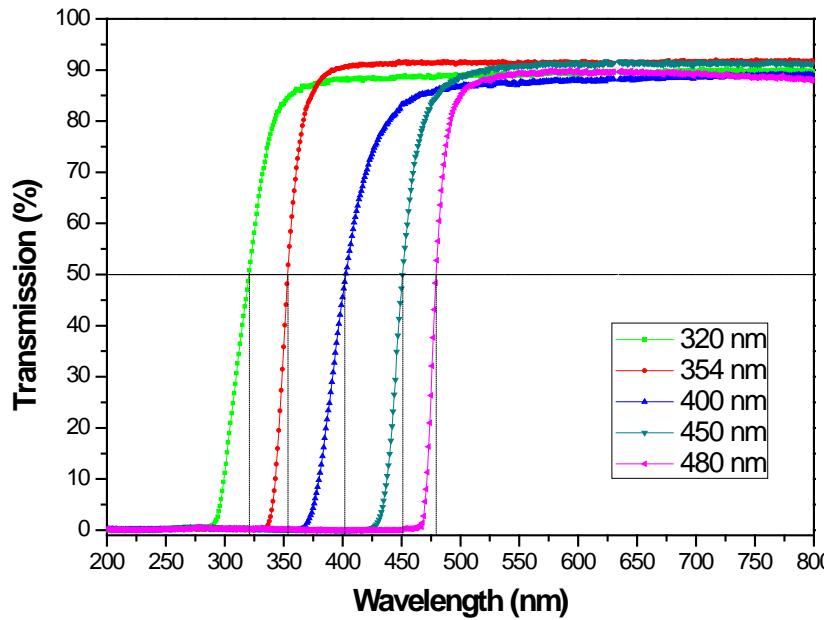

**Fig. 2** Transmission curves of the long wave pass filters used in this work. The designated wavelength for each filter means that the light transmittance is 50% at the wavelength.

Moreover, the activity of Cu-doped ZnS increased slightly as the amount of Cu dopant increased and reached a maximum value at x = 0.3% (Fig. 3). With further increase of x, the activity continuously decreased. This may be due to the color of the doped ZnS particles which acts as an optical filter and shortens the penetration distance of the incident light in the colloidal suspensions. On the other hand, at high doping concentrations, the possible appearance of a separate CuS phase may aggravate photoelectron-hole recombination in the volume of the photocatalysts (Wang et al. 2012b) as well as shielding the incident light, which results in the significant reduction of the photocatalytic activity.

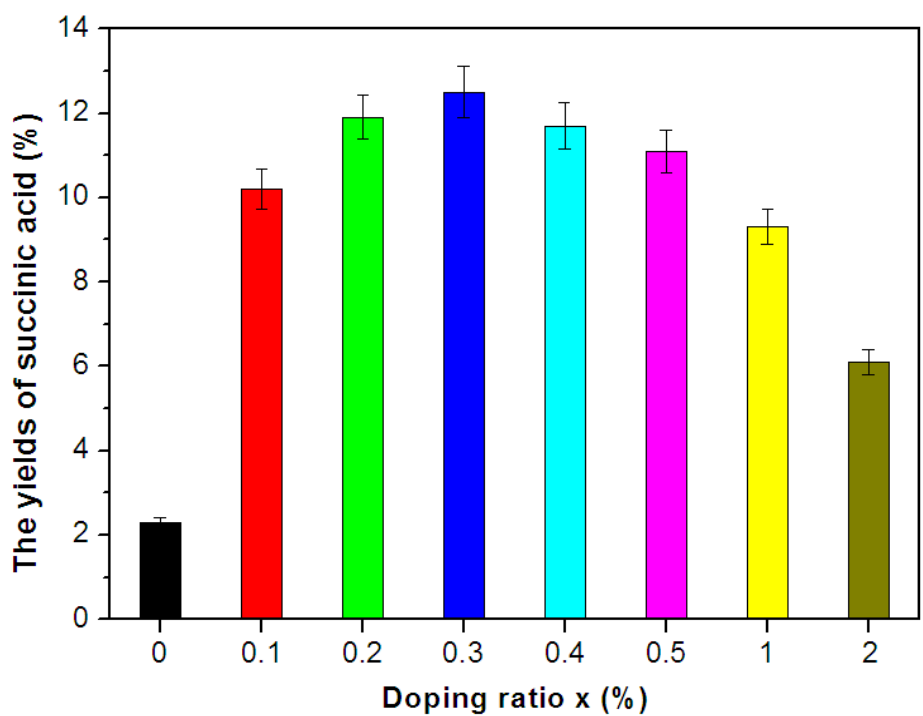

**Fig. 3** Comparison of the photocatalytic efficiencies of pure and Cu-doped ZnS prepared at different x. Reaction time: λ > 354 nm, 2 h. The error bars represent standard deviations from three independent experiments.

In further experiments, we examined the photocatalytic efficiencies of $Zn_{1-x}Cu_xS$ (x = 0.3%) under irradiation of five different wavelength ranges. Figure 4 represents the amount of succinic acid produced from the reduction of fumaric acid during the photoreaction as a function of irradiation time. It can be found that under different irradiation conditions, the stoichiometric yields of succinic acid increased nearly linearly with the reaction time. In the case of λ > 320 nm irradiation, they reached a constant level at 6 h (Fig. 4a). This may result from the exhaustion of the reactants (fumaric acid and $Na_2SO_3$) and also the undesirable oxidizing degradation of the product by photogenerated holes in the semiconductor. More interestingly, even light of λ > 450 nm can drive the photoreaction to produce micro-amounts of succinic acid.

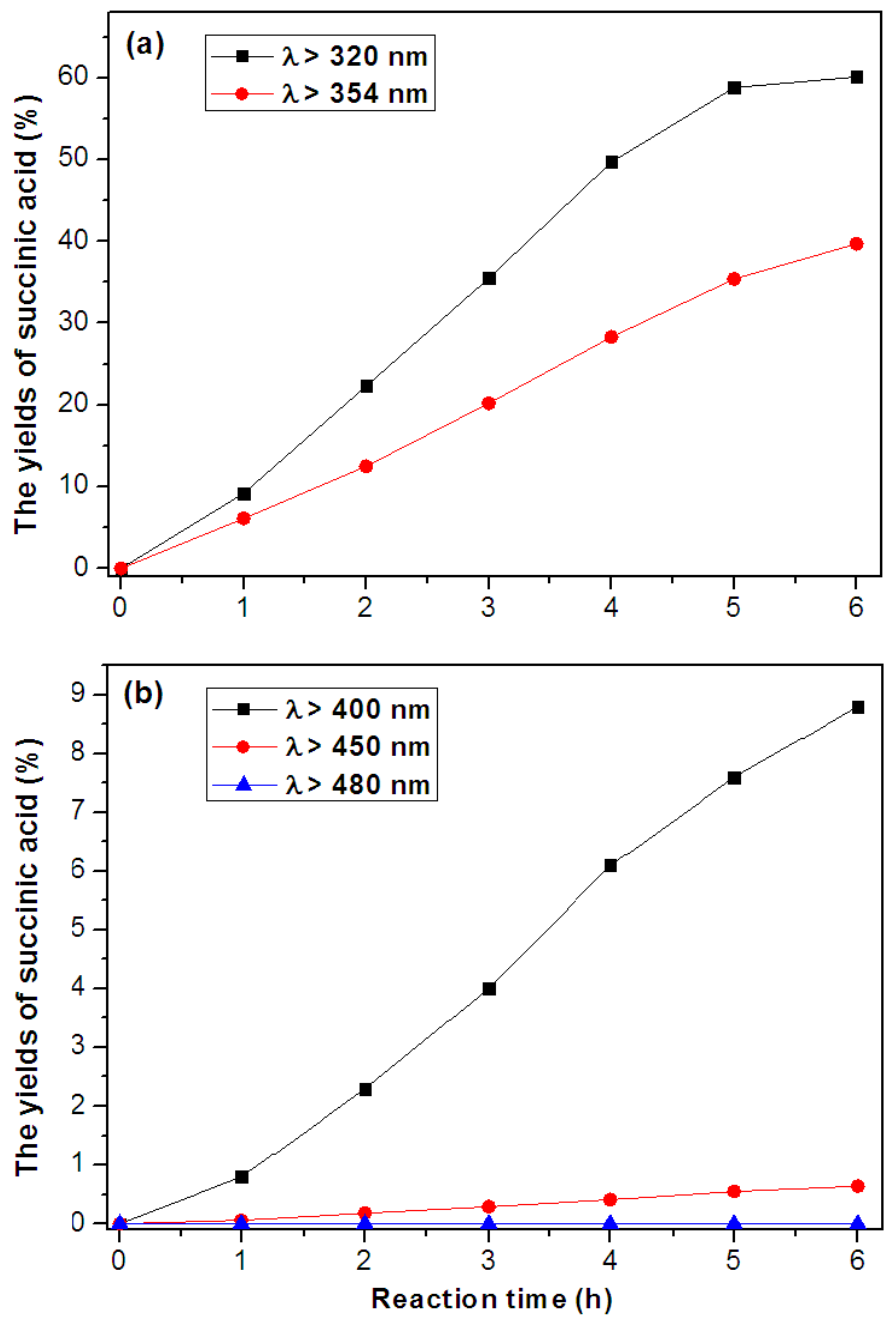

**Fig. 4** Time course of succinic acid production on photo-illuminated $Zn_{1-x}Cu_xS$ (x = 0.3%) surfaces, with light of five different wavelength ranges.

**Conclusions**

In this paper we investigated the influence of Fe, Ni, and Cu doping on the photocatalytic efficiencies of ZnS. The results show that the photocatalytic activity of ZnS does not increase by doping Fe and Ni, but can be significantly enhanced by doping Cu because this dopant can alter the electronic band structure of ZnS and make its optical absorption edges shift from UV band (344 nm) to longer wavelengths. The optimal doping concentration of Cu was found to be 0.3 atom%. Even under λ > 450 nm light irradiation, the photocatalyst $Zn_{1-x}Cu_xS$ can drive the reduction of fumaric acid to produce succinic acid. Given the existence of this doped semiconductor in the hydrothermal vents on early Earth and its capability to utilize both UV and visible light, ZnS might have participated more efficiently than previously estimated in the prebiotic chemical evolution.

**Acknowledgements** This research is supported by National Natural Science Foundation of China (No. 40902014) and Natural Scientific Research Innovation Foundation in Harbin Institute of Technology (HIT.NSRIF.2013055).